\documentclass[pra,twocolumn,notitlepage]{revtex4-1}

\usepackage{graphicx}
\usepackage{comment}
\usepackage{amsmath}
\usepackage{physics}
\usepackage{algorithmic}
\usepackage{qcircuit}

\begin{document}

\title{\Large{\textbf{Machine learning the period finding algorithm}}}
\author{John George Francis}
\affiliation{School of Physics, IISER Thiruvananthapuram, Kerala, India 695551}
\email{johnfrancis15@iisertvm.ac.in}

\author{Anil Shaji}
\affiliation{School of Physics, IISER Thiruvananthapuram, Kerala, India 695551}

\date{\today}

\begin{abstract}
    We use differentiable programming and gradient descent to find unitary matrices that can be used in the period finding algorithm to extract period information from the state of a quantum computer post application of the oracle. The standard procedure is to use the inverse quantum Fourier transform. Our findings suggest that that this is not the only unitary matrix appropriate for the period finding algorithm, There exist several unitary matrices that can affect out the same transformation and they are significantly different from each other as well. These unitary matrices can be learned by an algorithm. Neural networks can be applied to differentiate such unitary matrices from randomly generated ones indicating that these unitaries do have characteristic features that cannot otherwise be discerned easily. 
\end{abstract}

\maketitle
\newpage

\section{Introduction}
In the recent past machine learning has caught on as a viable tool in various fields and physics has been no exception. Design of Quantum algorithms has been, in general, a relatively slow process since the advent of quantum computation. Starting from the idea put forward by Feynman~\cite{feynman1982simulating} there are still only a few remarkable quantum algorithms that stand out distinctly. Deutsch~\cite{deutschjozsa_algorithm} was the among the first to show that there is an algorithm implementable on a quantum computer that runs in exponentially lesser oracle queries than its classical counterpart. Simon~\cite{simonsalgorithm} came up with another example of a problem and a quantum algorithm that runs exponentially faster. The major breakthrough came in 1994 when Shor came up with a quantum algorithm for a problem of great practical significance. An algorithm for integer factorisation in polynomial time, at the center of which sits the quantum period finding algorithm. The  algorithm  identifies the period of a given function, given an oracle which implements evaluation of a periodic function.
Classically, repeated function evaluation until the period is ascertained takes at least $r$ evaluations where $r$ is the period of the function. However, with a quantum oracle one can determine the period in just a few  evaluations of the oracle.

The general structure of the algorithm for oracle type of problems involves three stages. The first is a preparation stage where the input register of a quantum computer is put into a uniform superposition of all possible states. This facilitates access to all possible inputs on which  further computation may be done. The second stage is the oracle evaluation which evaluates the oracle on this superposition state. The oracle has a well defined action on every computational basis state of the input register. Its action on the superposition state therefore yields a superposition of the evaluations on every basis state. In essence, all evaluations on possible basis states are available in superposition after applying the oracle. In other words, the 'correct' answer to the computational problem is available at the output end of the oracle but it is hidden within a massive superposition of all possible outputs of the oracle. The third stage of the computation is a post processing that is also the clever bit of the algorithm. The post processing stage is where the information that the algorithm aims to determine about the oracle is extracted from the state after oracle evaluation. Design of the quantum post processing requires bringing together knowledge of hidden symmetries of the problem, deep number theoretic relationships, understanding of integral transforms and the ability to 'think quantum-mechanically' for implementing the desired transformations on a quantum register. The detailed example of the period finding algorithm, explained in Sec.~\ref{period}, further illustrates this point.

In the paper we investigate whether the quantum post-processing step can be designed with assistance from machine-learning. This is done for the specific case of the quantum period finding algorithm for which an implementable post processing step is already known that involves the inverse quantum Fourier transformation. The post-processing step boils down to a unitary transformation on the superposition of oracle outputs that leads to a state on which specific measurement can reveal the answer one is looking for. In our case, the period of the function is the answer that is sought. The objective of the machine learning scheme is to produce a post-processing unitary that puts the output register in quantum states such that the unknown period of the function can be deduced from its measurement statistics. The overall strategy is to compute the elements of this unitary post-processing transformation by minimizing a cost function that is constructed based on the desired measurement statistics. 

The input into the machine learning algorithm are all the elements of the post-processing unitary. The cost function has to be optimized with respect to these inputs which, in turn, are exponential in number with the respect to the number of input qubits on which the period finding algorithm is implemented. Differentiable programming is a paradigm of programming where the derivatives of the output (cost function) can be taken with respect to the large number of inputs at the cost of a small overhead using automatic differentiation~\cite{autodiff} (as opposed to finite difference or symbolic differentiation or coding the derivatives manually). This allows us to employ the gradient descent optimization described in detail in Sec.~\ref{gradient} for finding the matrix elements that minimize the cost function. In fact, a central part of many deep learning and other machine learning approaches is back-propagation which is a special case of differentiable programming. In this paper we use differentiable programming~\cite{differentiable_programming} to find the post processing unitary. Significantly we find that the post-processing unitary is not unique even through till date only one such unitary - the inverse quantum Fourier transform - is known. We also find that we are able to use neural networks to classify unitary matrices based on their usefulness as post-processing unitaries for the period finding algorithm, indicating that the unitaries produced by the machine learning algorithm have common features that may not be evident from inspection or from their spectral characteristics.

Machine learning approaches have previously been employed for the analysis and discovery of quantum algorithms. Gepp and Stocks \cite{Gepp2009} gives a review of the application of genetic algorithms to evolve quantum algorithms. Lukaz \textit{et al}~\cite{Cincio_2018} demonstrate a machine learning approach to discovering short-depth algorithms for reducing computational errors on near term quantum computers. Along the lines of the work presented in this Paper, Bang \textit{et al.}~\cite{Bang_2014} use a classical-quantum hybrid simulation along with differential evolution to arrive at unitary matrices for effecting out the Deutsch-Josza algorithm~\cite{deutschjozsa_algorithm}. Design of unitary matrices for Simon's algorithm is investigated in~\cite{wan2018learning}. In~\cite{mauro} the use of a variational algorithm to recover a quantum circuit for Grover's algorithm \cite{Groversalgorithm} is demonstrated.  Our work takes this a step further and considers a quantum algorithm that is of practical importance, namely the quantum period finding. We also address the what is typically the hardest part of the algorithm design, namely the post-processing. Assistance from machine learning in this step can potentially lead to implementable algorithms for the likes of the hidden subgroup problem for which an efficient quantum algorithm is believed to exist but is not known at present in the general case.

In the next section we outline the period finding algorithm. Sec.~\ref{ML} details the method used in implementing a computer program that finds a unitary matrix for the post oracle processing. The gradient descent, which is the optimisation procedure we use, is discussed in Sec.~\ref{gradient}. We also compare the measurement statistics of the output states produced by the post-processing unitaries generated by the gradient descent algorithm with that from the standard quantum period finding algorithm in this section. We classify the post-processing unitaries produced by machine learning algorithm using a feed-forward neural network in Sec.~\ref{unitary} and discuss the limitations of our approach in Sec.~\ref{training}.  We summarize our results in section~\ref{conclusion}.
 
\section{The period finding algorithm \label{period}}

A function that takes $n$-bit numbers (integers) as input and outputs $n$-bit numbers of the form 
\[ f : \{0,2,3,4,...2^n-1\} \rightarrow \{0,2,3,...2^n-1\},\] 
is given. The function is known to be periodic with an unknown period $r$ i.e $f(x + r) = f(x)$. In addition, it is stipulated that the values of $f$ do not repeat within each period, i.e $f(i), f(i+1), f(i+2),... f(i+r-1)$ are all unique for all values of $i$. Since the function is from integers to integers, this condition excludes only very few periodic functions. More importantly, this class encompasses the types of functions encountered in order finding that, in turn, is used in the integer factoring algorithm. Given the task of determining the period $r$, one approach is to compute $f(1), f(2)... f(i)$ until we get a value $f(s)$ equal to $f(1)$. Then $s=r$ will give the period of the function. However this approach requires at-least $r$ function evaluations and $r$ can be exponential in $n$ since the domain of the function is also exponential in $n$ ($r$ could be as big as $2^{n-1}$). This simple and straightforward approach whose computational complexity is exponential in $n$ also turns out to be the best known deterministic classical algorithm to find the period. 

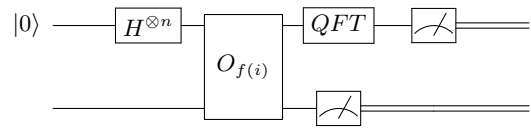
\begin{figure}[htbp!]
	\mbox{
\Qcircuit @C=0.7em @R=0.2em @! {
	& \lstick{\ket{0}}  & \gate{H^{\otimes n}} & \multigate{1}{ O_{f(i)}} & \gate{QFT} &\meter & \cw \\
	& \lstick{} & \qw  & \ghost{O_{f(i)}} &  \meter & \cw & \cw
} }	
	\caption{Quantum circuit for the period finding algorithm \label{period1}}
\end{figure}

The quantum period finding algorithm evaluates the period of a function in queries polynomial in $n$. We briefly recap the algorithm for completeness and for establishing the notation we use. The quantum circuit for the period finding algorithm is given in Fig.~\ref{period1}.

The circuit consists of a register $X$ of $n$ qubits with Hilbert space of dimension $N = 2^n$ and another register $F$ with $m$ qubits. All the qubits in both registers $X$ and $F$ are initialized into the state $\ket{0}$. In the first step of the algorithm, Hadamard gates are applied to each qubit in the register $X$. After these Hadamard gates the joint state of the two registers registers is 
\[ |\psi^{(1)}\rangle_{XF} = \frac{1}{ \sqrt{2^n} } \sum_{i} \ket{i} |\mathbf{0} \rangle, \quad i = 0, \ldots, 2^{n}-1 .\]
The oracle is a unitary transformation acting on the two registers $X$ and $F$ which implements the function in question. It takes the basis state $\ket{i}\ket{0}$ to $\ket{i}\ket{f(i)}$. Thus its action on $\ket{\psi^{(1)}}_{XF}$ is to produce the state
\[ |\psi^{(2)}\rangle_{XF} = \frac{1}{\sqrt{2^n}} \sum_{i} \ket{i} \ket{f(i)}.\]
A measurement on the second register ($F$) that yields a value $f(i_0)$ with probability $1/r$ collapses $|\psi^{(2)}\rangle_{XF}$ to
\begin{equation} 
    \label{eq:psi3}
|\psi^{(3)} \rangle_{XF} = \frac{1}{\sqrt{[2^n/r]}}  \sum_{p=0}^{[2^n/r] - 1} \ket{i_0 + pr}\ket{f(i_0)}.
\end{equation} 
The (inverse) quantum Fourier transform is then applied on the first register as implemented by the corresponding post-processing unitary matrix which is the discrete Fourier transform matrix. Using
\[ \sqrt{\frac{N}{r}} \sum_{p=0}^{N/r -1}\ket{i_0 + pr}  \xrightarrow[]{QFT} \frac{1}{\sqrt{r}} \sum_{q=0}^{r-1} \varphi_{i_0}^q| qN/r\rangle, \]
where $\varphi_{i_0}^q$ is a phase dependent on the offset $i_0$ and $q$ that is not important for our discussion, we obtain the state of the two registers after the Fourier transformation as 
\begin{equation}
    |\psi^{(4)}\rangle_{XF} = \frac{1}{\sqrt{r}}   \sum_{q=0}^{r-1}  \ket{q[2^n/r]}\ket{f(i_0)}. 
\end{equation}
Measuring the first register in the computational basis yields any one of the $r$ possible values, $q2^n/r$ with probability $1/r$ respectively. The measurement statistics obtained by projecting the register $X$ on to the complete set of computational basis states is therefore a probability distribution with equal height peaks at $2^n$/r intervals if $2^n$/r is an integer. If $2^n$/r is not an integer, then the distribution has narrow peaks around the integer $[2^n/r]$. By sampling this distribution, finding the location of these peaks and with some classical post processing~\cite{cont_frac}, to any desired level of accuracy one can compute the period, $r$, of the function. 

\begin{figure}[!ht]
    \resizebox{8.5 cm}{5.5 cm}{\includegraphics{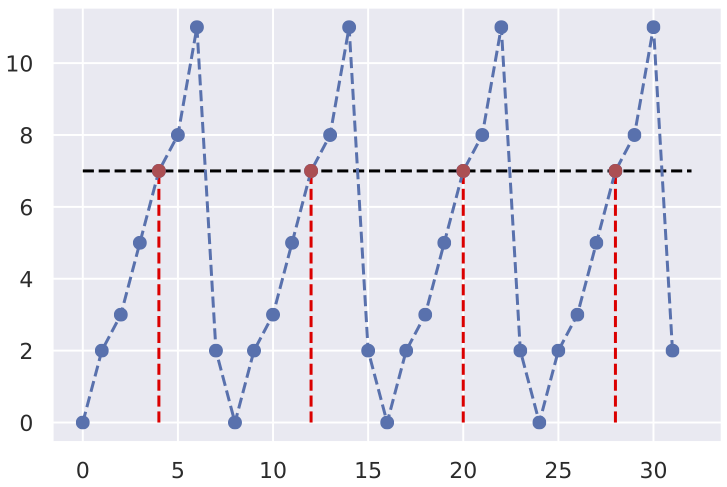}}
	\caption{Example function with period 8. If the measurement on the second register yields the answer 7 after the oracle, then the amplitudes of the first register are concentrated on states $\ket{j}$, where $j$ are marked by the red lines in the figure.  \label{dist}} 
\end{figure}

For example, suppose we had a function of period 8 and register $X$ had 5 qubits. After the oracle, if the measured value on the second register was $7$, corresponding to  $\ket{7}$, the state of the system would be 
\[ |\psi^{(4)} \rangle_{XF} = \frac{1}{2} \sum_{t=0}^3  \ket{7+8t} \ket{7}. \]
Note that the amplitudes on the first register are concentrated on $\ket{j}$ where $j$ correspond to the red lines in Fig.~\ref{dist}. The periodicity of these amplitudes is revealed by the final post-processing step, namely the Fourier transform. It can be easily seen that the structure of the state of the $X$ register in $|\psi^{(3)} \rangle_{XF}$  is independent of the outcome of the first measurement on the register $F$. We can therefore just as well ignore this measurement result which would be equivalent to tracing over the register $F$ after the action of the post-processing unitary. 

The period finding problem is a hidden-subgroup problem (HSP) on the group $Z_n$. The standard way to approach Abelian HSP on a quantum computer is to put the system first in a uniform superposition of all possible input states and then apply the oracle. This puts the first register of the system in a state of uniform superposition of cosets of the hidden subgroup~\cite{hsp_lomont}. Revealing the hidden subgroup from this state is achieved  by an appropriate Fourier sampling. However for Non Abelian groups Fourier sampling cannot reveal the information about the hidden subgroup and no general algorithm is known for arbitrary non-abelian groups. Machine learning approaches like the one we  use potentially be used to reveal unitary matrices for the quantum post-processing for the case of Non Abelian subgroups and solve the HSP. This could also, in principle, be used to show that there exists unitaries that can reveal hidden subgroups albeit for small groups due to the prohibitive computational cost of simulating quantum computations on classical computers.   

We show here that using gradient descent learning one can design a unitary matrix that carries out the quantum post oracle processing which reveals the period of the function. The unitary that is obtained is typically not identical to the quantum Fourier transform unitary. This shows that there exists other unitaries that effect out the same transformation. Gammelmark and Molmer~\cite{Gammelmark_2009} have previously worked on the quantum Fourier transform and used machine learning to improve the approximations to the transformation whose experimental implementation is simpler. The quantum Fourier transform implemented as gates is a series of controlled phase shifts of the form $\pi/2^m$ that connect all the qubits in a register to one another.  Even if the number of two qubit gates required in this case scale  as a polynomial in $N$, in practice, gates connecting each qubit to all other qubits are difficult to implement. The question addressed using machine learning in~\cite{Gammelmark_2009} is whether non-standard phase shifts can approximate the Fourier transform better than the usual fixed phase shifts if the connectivity between qubits is restricted. The investigations in~\cite{Gammelmark_2009} leads to the conclusion that indeed this is the case. In view of this result, it is not entirely surprising that there exists other unitary matrices that can carry out the same post-processing for extracting period information.

\section{Machine learning the unitary matrix \label{ML}}

The objective here is for a computer program to find a unitary matrix that does the post processing and reveals the period using stochastic gradient descent or a similar optimizer. Stochastic gradient descent is  a method frequently used in machine learning to train neural networks. The program simulates the three stages of the quantum period finding algorithm. The unitary matrices in the first two stages that put register $X$ in a uniform superposition and then perform the function evaluation (oracle) are kept fixed in the simulation. The program is then tasked with finding the third stage post-processing unitary matrix which generates the output state of the algorithm on which a straightforward measurement will reveal the period of the function.

We have a set of functions, $\{f_j\}$, with known periods that use as the training data. The periods of the functions are limited to less than half the size of the domains of the functions. For a given function  $f_j$ from the training data a corresponding program function to implement $U_2$ is generated. $U_2$ is the oracle unitary operator which is defined by $U_2 (\alpha \ket{i}) \rightarrow \alpha \ket{i} \ket{f(i)} $ and it depends on the function $f_j$ from the training data set that is used. The post-processing unitary matrix that we task the program to find is labeled as $M_3$. The matrix $M_3$ is initially chosen to be an arbitrary (not necessarily unitary) matrix that converges to a unitary matrix with the desired property as the program progresses in its iterations. The number of real independent parameters for an arbitrary $ m \times m$ complex matrix is $2 m^2$. An $n$ qubit system has operators of size $2^n \times 2^n$, and therefore the number of real independent parameters is $2\times2^{2n}$.  

These $2 \times {2^{2n}}$ parameters will be tuned in the optimization to get the desired unitary matrix. The $X$ register is initialized with $\ket{0}$ and transformed through the three transformations $U_1 = H^{\otimes n}$, $U_2(f_i)$ and $M_3$. The final state is obtained and the probability distribution $P(i)$ for getting integer $i$ on measurement of register $X$ is calculated from this state after tracing over the second register as mentioned earlier. This means that the measurement operator on the second register need not be implemented in the simulation. Out of the final state of the two registers all we have to do is to pick out the probabilities corresponding to each computational basis state of the first register and sum over the associated states of the second. So, in order to get the probability distribution from the final state $|\tilde{\psi}^{(4)} \rangle$ (the tilde denoting the suppressed intermediate measurement), we write the state in the computational basis as 
\[ |\tilde{\psi}^{(4)} \rangle  = \sum_{i,j} \alpha_{ij} \ket{i,j} \]
The Probability of getting the output $i$ on measuring register $X$ is, \[ P(i) = \sum_{j=0}^{r-1} \left| \alpha_{ij} \right|^{2} \]
In other words the probability of getting $i$ on measurement is the sum of the absolute value squares of the amplitudes of the states whose first $2^n$ bits in binary are a binary representation of $i$. 

The loss function whose  value will be minimized during the training(optimization) phase is    
\begin{equation} \label{loss} 
g_{\rm{loss}} = \frac{1}{2^n}\sum_{i} \big[ P_a(i) - P_d(i) \big] ^2 + \frac{k}{2^{2n}}\sum_{i,j} | M_3^\dagger M_3 - I  |^{2}_{ij}.
\end{equation}
Here $P_a$ (actual probability distribution) is the probability distribution of the $X$ register from the training circuit and $P_d$ (desired probability distribution) is the  probability distribution of $X$ register from the conventional period finding circuit. The first sum in $g_{\rm{loss}}$ is a measure of the how close the distributions $P_a$ and $P_d$ are. It is always positive and goes to zero only if the distributions are exactly identical. The second  term is a unitary penalty term and it quantifies how far the matrix $M_3$ is from being a unitary matrix. It is also always a positive number and  goes to zero only if $M$ is a unitary matrix. Thus when $g_{\rm{loss}}$ is zero or close to zero, the distributions $P_a$ and $P_d$ are identical or close to identical and the operator $M_3$ is unitary/almost unitary. 

The loss function, $g_{\rm{loss}}$ depends on the $2 \times 2^{2n}$ variables that are elements of $M_3$ for a given function $f_j$ from the training data set. The loss function has to be minimized with respect to the elements of $M_3$ across all functions in the training data set. For doing the minimization we use a variant of the gradient descent algorithm called ADAM (Adaptive Moment Estimation)~\cite{adam} which is a popular choice in the field of machine learning to training neural networks.  
\begin{figure}[!ht]
	\centering
	\resizebox{8.5cm}{5cm}{\includegraphics{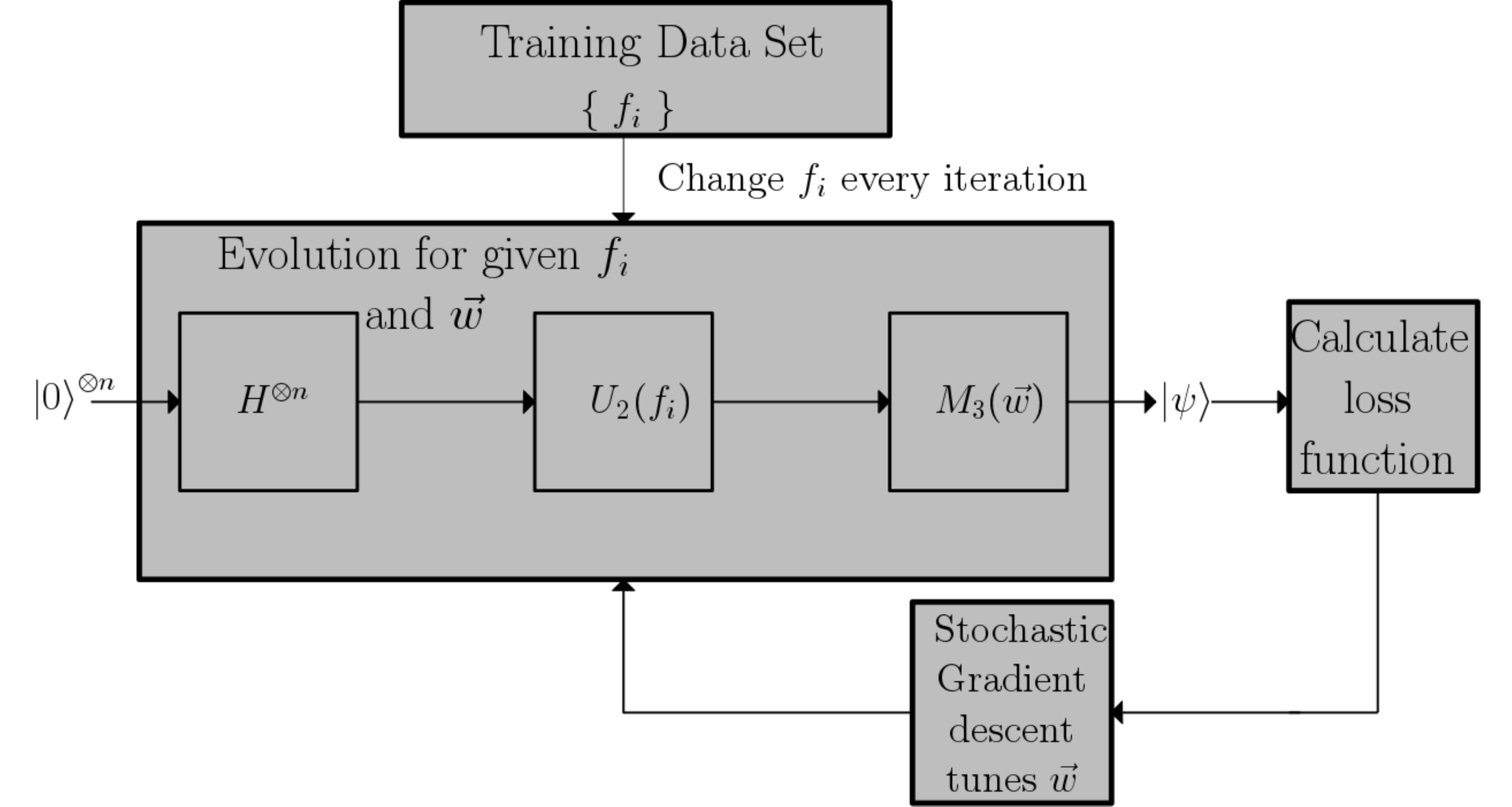}}
	\caption{Schematic representation of the optimisation procedure}
\end{figure}

\section{Gradient Descent optimisation \label{gradient}}

Gradient descent is a procedure to find the minimum of a multi-variable function. Typically we have a function (over a data set) that we want to minimize:  
\[ G(\vec{w}) = \frac{1}{n}\sum_{j}^{n} g_j(\vec{w}),\]
where each $g_j(\vec{w})$ corresponds to an instance of the training data. $G(\vec{w})$ is therefore the average over the training data. At each iteration $\vec{w}$ is updated by the rule, 
\[ \vec{w} :=  \vec{w} - \alpha \grad g_j(\vec{w}) \]
where $\alpha$ is step size, also called the learning rate. This iteration is performed through for all $j$, namely all samples in the training set. This constitutes one epoch of training. Multiple epochs are repeated till the function $G(\vec w)$ converges to a desired minimum value. Selection of the step size (or learning rate), $\alpha$, is important. A low value for $\alpha$ makes training reliable but it takes longer for the training to converge on the minimum. On the other hand, large $\alpha$ values may lead to the training not converging or even diverging. This method of iterating and updating over every data point is called Stochastic Gradient Descent.

There exist several extensions and variants of the standard gradient descent that incorporates tuning the learning rate, adding ``momentum'' per parameter tuning rate etc.~\cite{gd_overview}. A popular choice is ADAM~\cite{adam}, which takes into account the first moment and the second moment of the gradients. Steps for ADAM  optimisation are listed in Appendix A in pseudo-code form. ADAM was found to converge the cost function to 0 faster than gradient descent and was the chosen method for all the optimisations in the following. ADAM takes 3 configuration parameters for the algorithm, namely the learning rate $\alpha$, and the exponential decay rates for moment estimates $\beta_1$ and $\beta_2$. We chose $\alpha$ to be 0.001 while $\beta_1$ and $\beta_2$ were set to 0.9 and 0.99 respectively. The functions $g_1$, $ g_2$, $g_3$... are the cost function values evaluated for different data points in the training data set. In our use case, elements of $M_3$ form $\vec{w}$ and $i$ indexes the training data is a set functions $f_i)$ of various known periods. Note that $g_{\rm{loss}}$ not only depends on $\vec{w}$ but also the training data-point it is evaluated on, i.e the periodic function $f_i$ defining the unitary matrix $U_2$. The function $g_{\rm{loss}}(\vec w)$ is evaluated and minimisation updates applied at a different data-points in every iteration of the loop, cycling through all the training data points till $G(\vec w)$ converges to the desired value, 0.

The optimisation procedure for 1 epoch on 5, 6 and 7 qubits took 0.343, 4.41 and 130.8 seconds respectively on a 28 core Intel Xeon Gold $6132$ cpu. We used TensorFlow\cite{tensorflow2015-whitepaper} (version 1.13) for its differentiable programming features and its optimizers. Higher qubit cases was found to converge faster. Cost function was found to reduce to a sufficiently small value ($\sim10^{-8}$) after 3000 epochs in the case of 5 qubits and 6 qubits and 2000  epochs in the case of 7 qubits. One run of the program compiles the TensorFlow computational graph, generates the dataset, runs the optimisation and saves the generated unitary to file. Time required and RAM usage for each run of the program was 17.71 minutes (1.0 GB), 222.4 minutes (3.8 GB) and 4363.85 minutes (20.2 GB) for the 5, 6 and 7 qubit cases respectively.

\subsection{Results of the optimization }

The program was run and training performed for 5, 6 and 7 qubit sized $X$ registers.  Given a size of register $n$, the  domain of the function is $\{0,1,2,3,4,... 2^n-1\}$. Periodic functions were generated with period randomly selected but which is less than half the domain size (16 for $n=5$, 32 for $n=6$ and $64$ for $n=7$). Data set size was 10, 15 and 20 for $n=5$, $6 $ and $7$ respectively. Training was run for 3000 epochs for 5 and 6 qubit cases and 2000 for 7 qubits,  which brought the average loss function over the data set to the order of $10^{-8}$. Several independent runs of the training were done and each run converged on a different set of $2^{2n}$ parameters (and thus a different unitary matrix) indicating significantly that there exists several other operators other than the inverse quantum Fourier transform unitary matrix prescribed in the standard period finding algorithm that effect out the same task. \begin{figure}[!htb]
\resizebox{8.5cm}{5cm}{\includegraphics{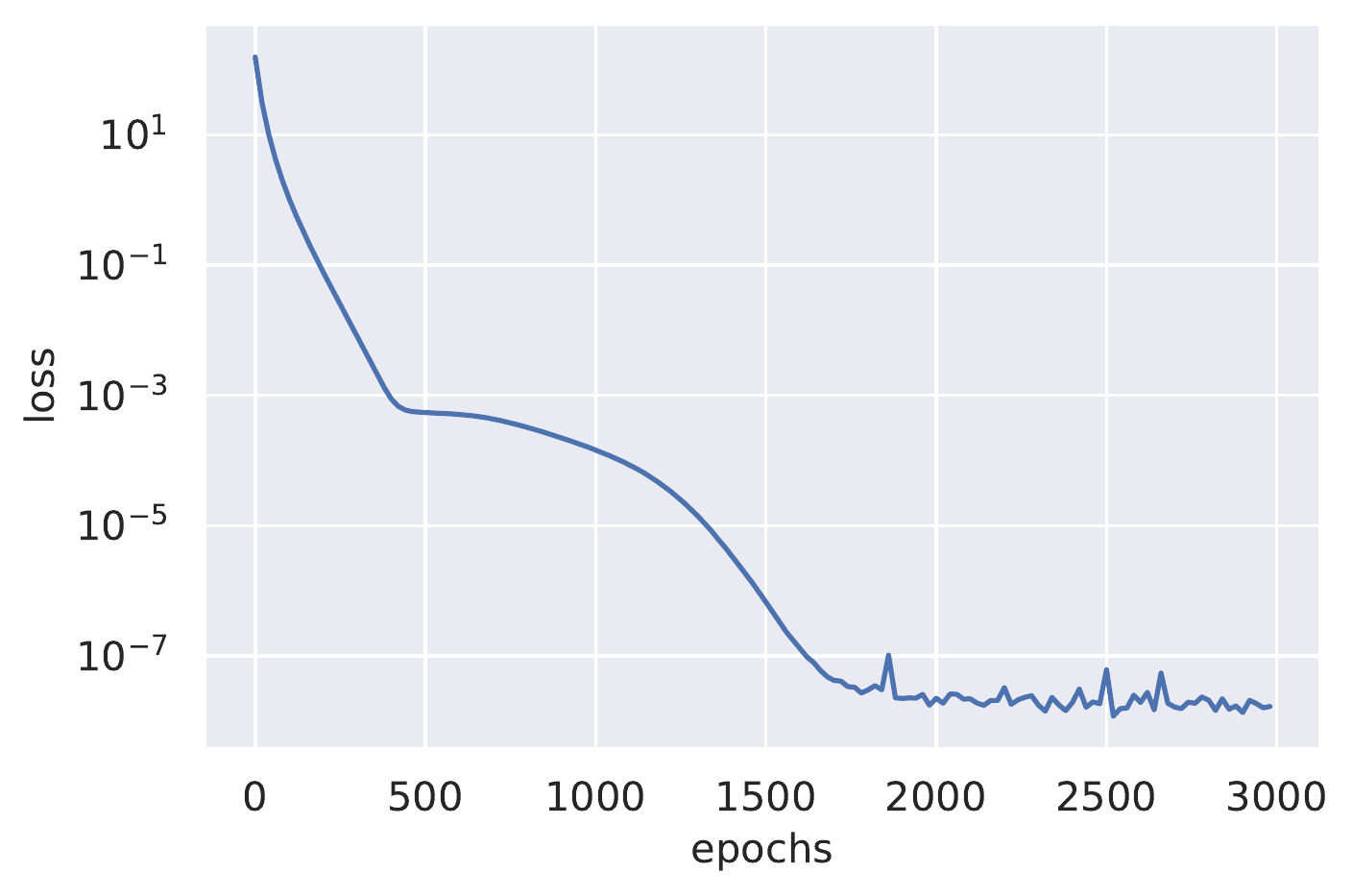}}
	\caption{The cost function as training progresses}
\end{figure}
%

In order to show that the distinct post-processing unitaries produced by the optimization are indeed differnt from each other in non-trivial ways, we used the Loschmidt echo. The Loschmidt echo~\cite{loschmidt_rodolfo} of two unitary matrices over a state quantifies the difference between the two unitary matrices when they act on that state. The Loschmidt echo of two operators $U_1$ and $U_2$ with respect a state $\ket{\psi}$ is.
\[ L = \left| \bra{\psi}U_1^{\dagger}U_2\ket{\psi}\right|^2.\]
Loschmidt echo ranges from 0 to 1 and a Loschmidt echo of 1 on a particular state signifies the unitary matrices have identical action on that state. The Loschmidt echo for the different unitaries obtained from our procedure were calculated against the inverse quantum fourier transform unitary matrix for the $\ket{\psi_0}_X = \ket{0}^{\otimes n}$. Loschmidt echo on this state were, in general,  much smaller than $1$. However Loschmidt echo on the state $H^{\otimes n} \ket{0}$ were close to $1$, indicating all these matrices have the same effect on the input state of the period finding algorithm, but are different otherwise. The Loschmidt echo over the uniform superposition state and the $\ket{0}^{\otimes n}$ state of some of the unitaries the program converged on are tabulated below, 
\begin{center}
	\begin{tabular}{| c | c| c| c|}
		\hline
		 & \vtop{\hbox{\strut Loschmidt echo on}\hbox{\strut uniform superposition}} & Loschmidt echo on $\ket{0}^{\otimes n}$  \\
		\hline
		1 & 0.99928572 & 0.00239334\\
		2 & 1.00008533& 0.00803578\\
		3 & 0.99900781& 0.01292283\\
		4 & 1.0035981 & 0.00297156\\
		5 & 0.99926629 & 0.0106097\\

		\hline

	\end{tabular}

\end{center}

A test data set with periodic functions of all period values was generated, including functions with periods not in the training data set. The unitary matrix the program converged on has an average value for the cost function of the order of $10^{-8}$ on the test data set. The Loschmidt echos on $\ket{0}^{\otimes n}$ were also much less than 1. The functions in the test data were not restricted to those having periods within half the domain size. Even for these, we find that the $M_3$ converged upon by the program has a small ($\sim 10^{-8}$) value for the cost function. Thus the program was able to find a unitary matrix that carried out the post-processing part of the period finding algorithm that worked on a wide variety of periodic functions on which it was tested.

\begin{figure}[!ht]
\resizebox{8.5cm}{12.5cm}{\includegraphics{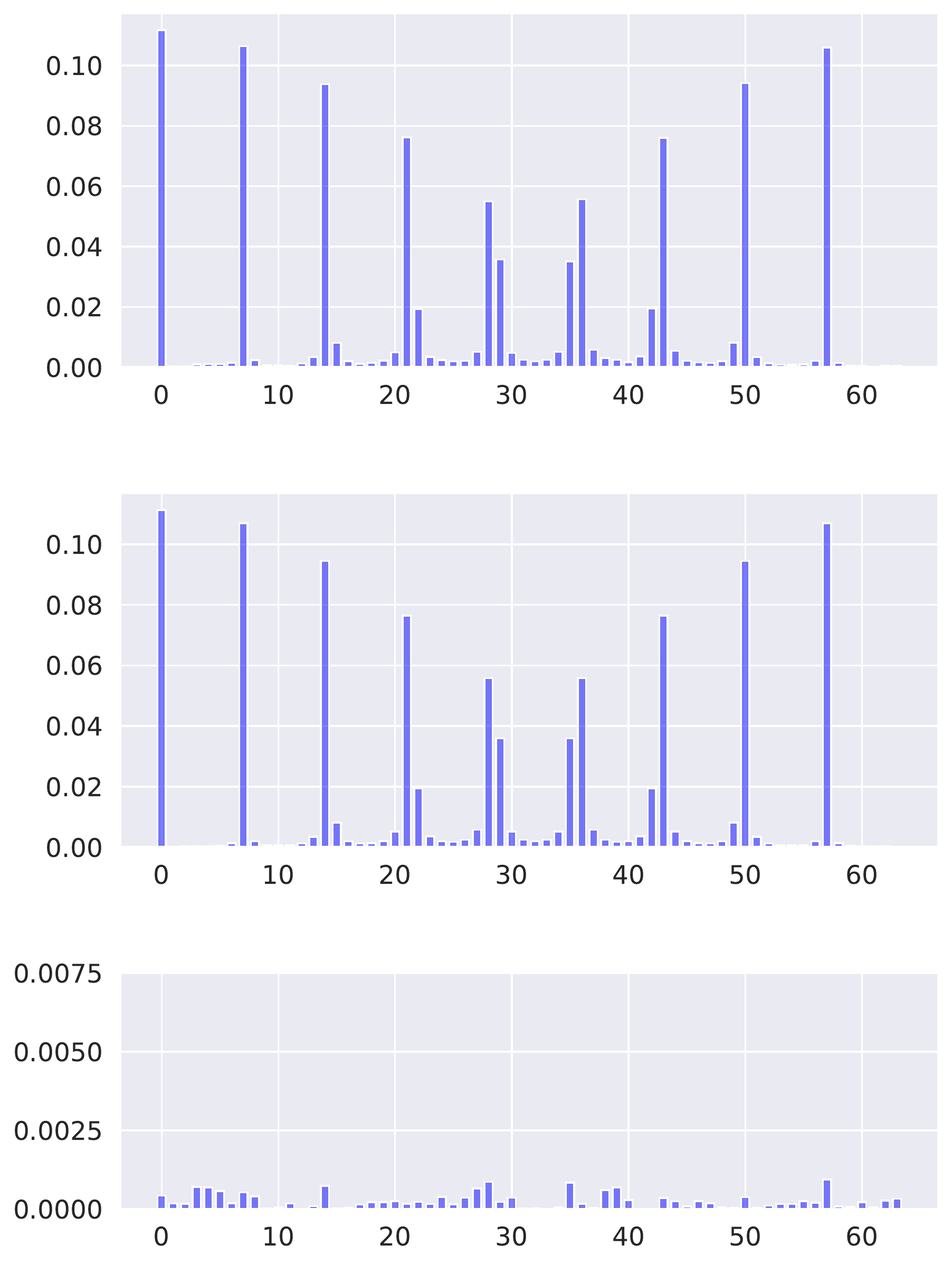}}
\caption{The figure on top shows the output probability distribution for a period 8 function when the quantum post-processing is done using the Fourier transform. The figure in the middle shows the same when the post-processing unitary obtained from the machine learning procedure is used. The figure in the bottom shows the difference between the two which is seen to be quite small. \label{compare}}
\end{figure}

In Fig.~\ref{compare}, it can be seen that the $M_3$ converged upon by the program gives a probability distribution very close to the distribution from the period finding algorithm as expected.  The plot on top is the probability distribution over $i$ of the target distribution that arises when the quantum Fourier transform is used on a period 8 function ($n=6$). The plot in the middle is the probability distribution when the post-processing unitary discovered by the algorithm is used and the one at the bottom is the absolute difference between the two distributions. We see that the absolute difference in probabilities are of the order of $10^{-3}$.

%
%

\begin{figure}
\resizebox{8.5cm}{5cm}{\includegraphics{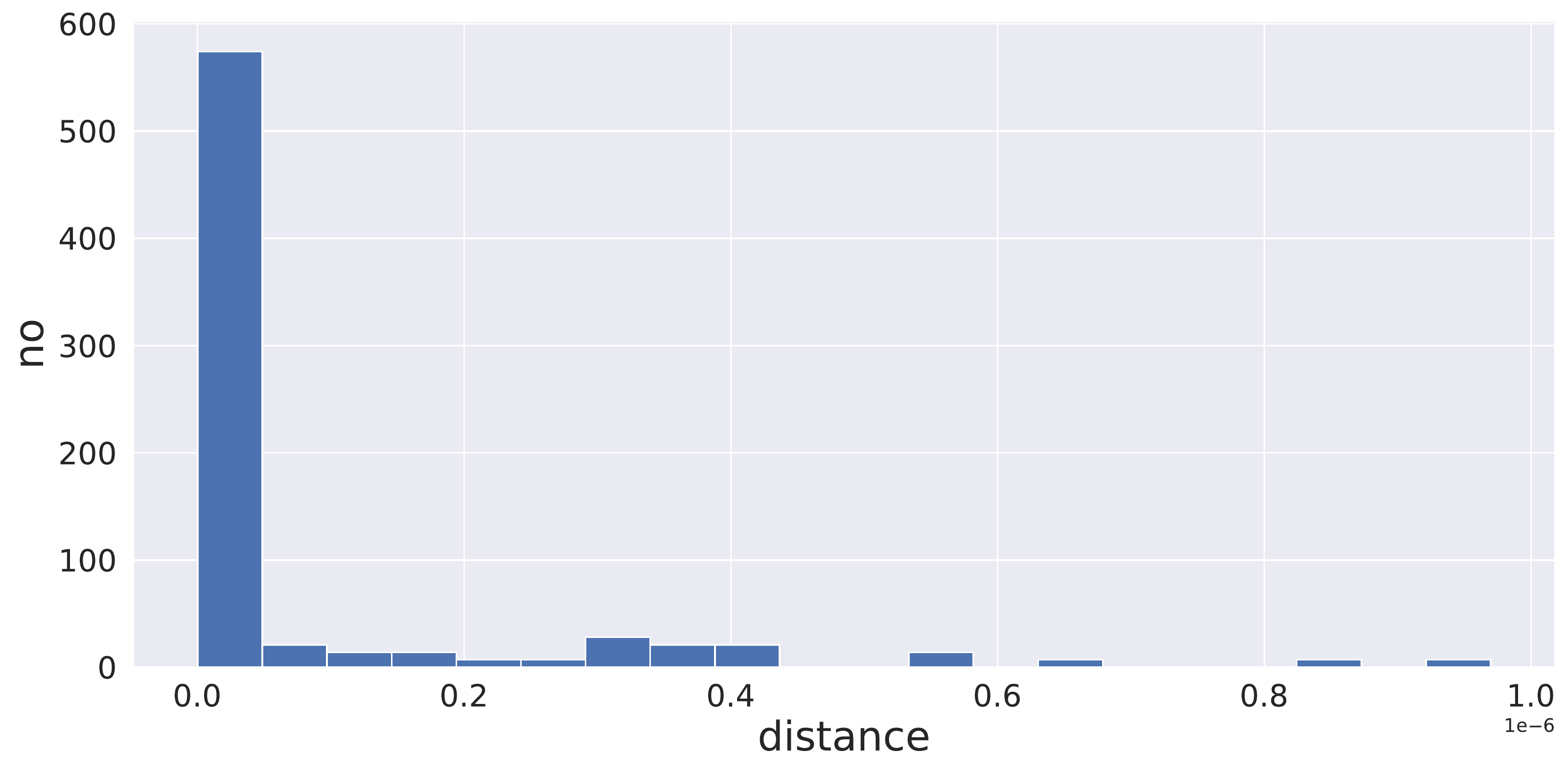}}
	\caption{Histogram showing distribution of distances between desired and obtained distributions of the various unitary matrices found  various periodic functions \label{dist1}}
\end{figure}

In order to see how closely the output statistics from the various unitaries produced by the algorithm compare with that produced by the Fourier transform, the following procedure was adopted. Unitaries for 5 qubit, 6 qubit, and 7 qubit cases were generated and the probability distributions obtained when these unitaries were used for post-processing were computed. For the same functions, the output distribution from the quantum Fourier transform were also generated. The comparison is done using a distance measure in the space of probability distributions,  ${\sum (p_i - q_i)^2}/{\text{len(p)}}$, where len(p) is the size of the sample space. It is found that the probability distributions arising when the generated unitaries are used are very close to that generated by the Fourier transform. Fig.~\ref{dist1} shows a histogram of the distribution of these distances for different functions of various periods of the unitaries (7 in each case) that were generated for 5, 6 and 7 qubit cases. The histogram is sharply peaked around 0, indicating that the two distributions that are being compared are virtually identical.

\section{Unitary Matrix classification using  Neural Networks \label{unitary}}

A natural question to ask is that if the unitary operators found by this algorithm have some pattern common to all of them. There is of course, the obvious similarity that they all produce output probability distributions that are virtually identical to the one produced by the Fourier transform. However as the analysis of the Loschmidt echo shows, this feature is particular to the action of the unitary when the equal superposition state is the input to the algorithm. Rather than depending on the finding similarities between the action of the unitaries on particular initial states, a deeper questions would be whether an analysis of the unitaries themselves would reveal a common pattern. 

\begin{figure}[!ht]
\resizebox{8.5cm}{5cm}{\includegraphics{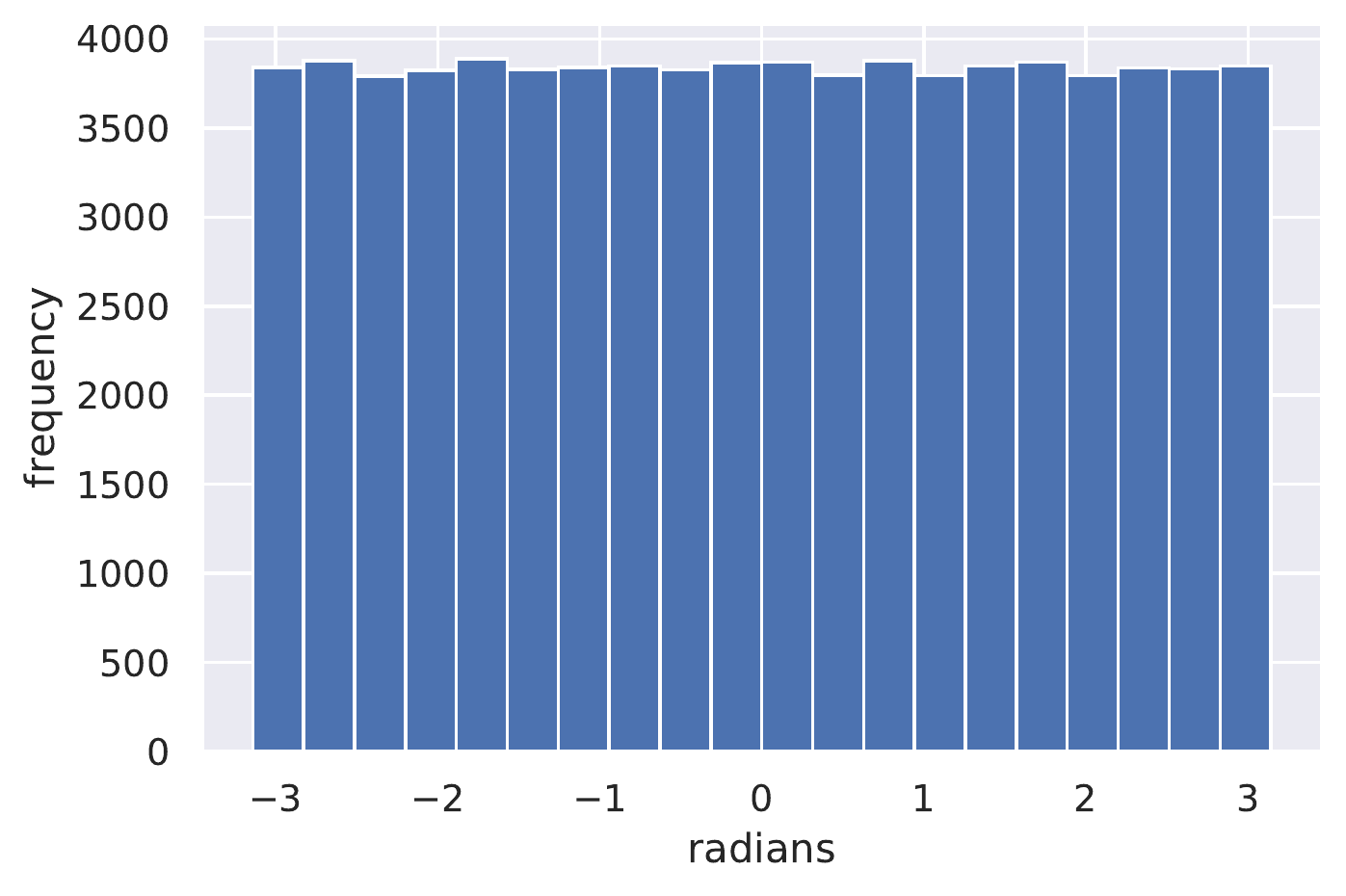}}
	\caption{Distribution of eigenphases of the post-processing unitary matrices generated by the machine learning procedure. \label{eigen1}}
\end{figure}

\begin{figure}[!ht]
\resizebox{8.5cm}{5cm}{\includegraphics{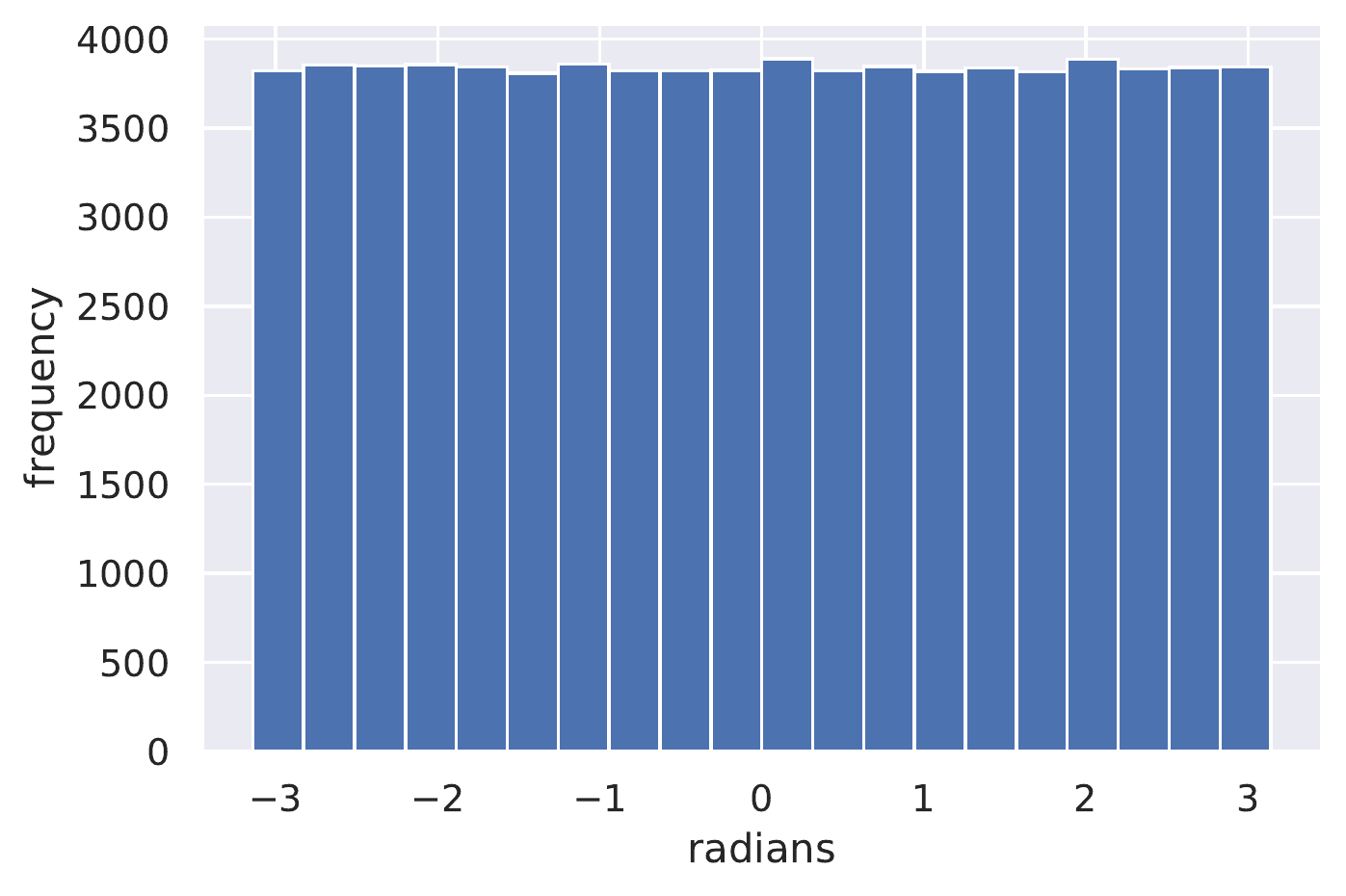}}
	\caption{Distribution of phases of randomly generated unitary matrices \label{eigen2}}
\end{figure}

Direct inspection of these relatively large unitary matrices did not lead to any similarities between them that are readily identifiable. As the next step we attempted to see if there are any similarities in the spectral features of the unitaries. A histogram of the distribution of eigenphases of the unitaries yields more or less a uniform distribution across 20 bins ranging form $-\pi$ to $\pi$ as shown in Fig.~\ref{eigen1}. 
A sample of randomly generated unitary matrices also show a similar distribution of eigenphases as seen from Fig.~\ref{eigen2}. So there appears to be no discernible pattern in the spectrum of the generated post-processing unitaries that distinguishes them a set of randomly generated ones.

Neural networks~\cite{neural} are machine learning tools that can learn complex patterns in data and can be then used to classify new data based on its learning. Neural networks are universal function approximators \cite{UFA_nn}. A data set can be thought of as a function from a set of inputs to outputs and a neural network learns this function during its training phase on the data set. To see if a neural network can learn a common pattern among these unitaries we trained it to classify the unitaries into two classes. Those of the type that can be used for quantum post-processing in the period finding algorithm and those that cannot. The full data set consists  of unitary matrices generated by the machine learning algorithm for 6 qubit systems as well as randomly generated unitaries both in equal numbers. 1200 of each were generated, thus a total of 2400 matrices. The  unitary matrices output by the algorithm have an associated value of $1$ and the randomly generated unitary has an associated value of $0$.  Out of this data set, 1800 unitaries were taken to train the neural network. A small portion (10\%) of the training data set is separated out and used as the validation data set. This set is not used for training the network parameters, but the loss and accuracy of the network evaluated on the validation data set is tracked during the training phase to check for over-fitting. The remaining 600 constitute the test data set for testing the predictions of the neural network after the training is done. Initially the loss and prediction accuracy of the training and validation data set improve continuously. After a point however, the loss and accuracy of the validation data set become poorer (higher loss and lower accuracy) while the loss and accuracy on the training data set keeps improving. At this point the neural network is considered to be over-fitting on the training data set and the training is stopped.

\begin{figure}
    \resizebox{8.5cm}{5cm}{\includegraphics{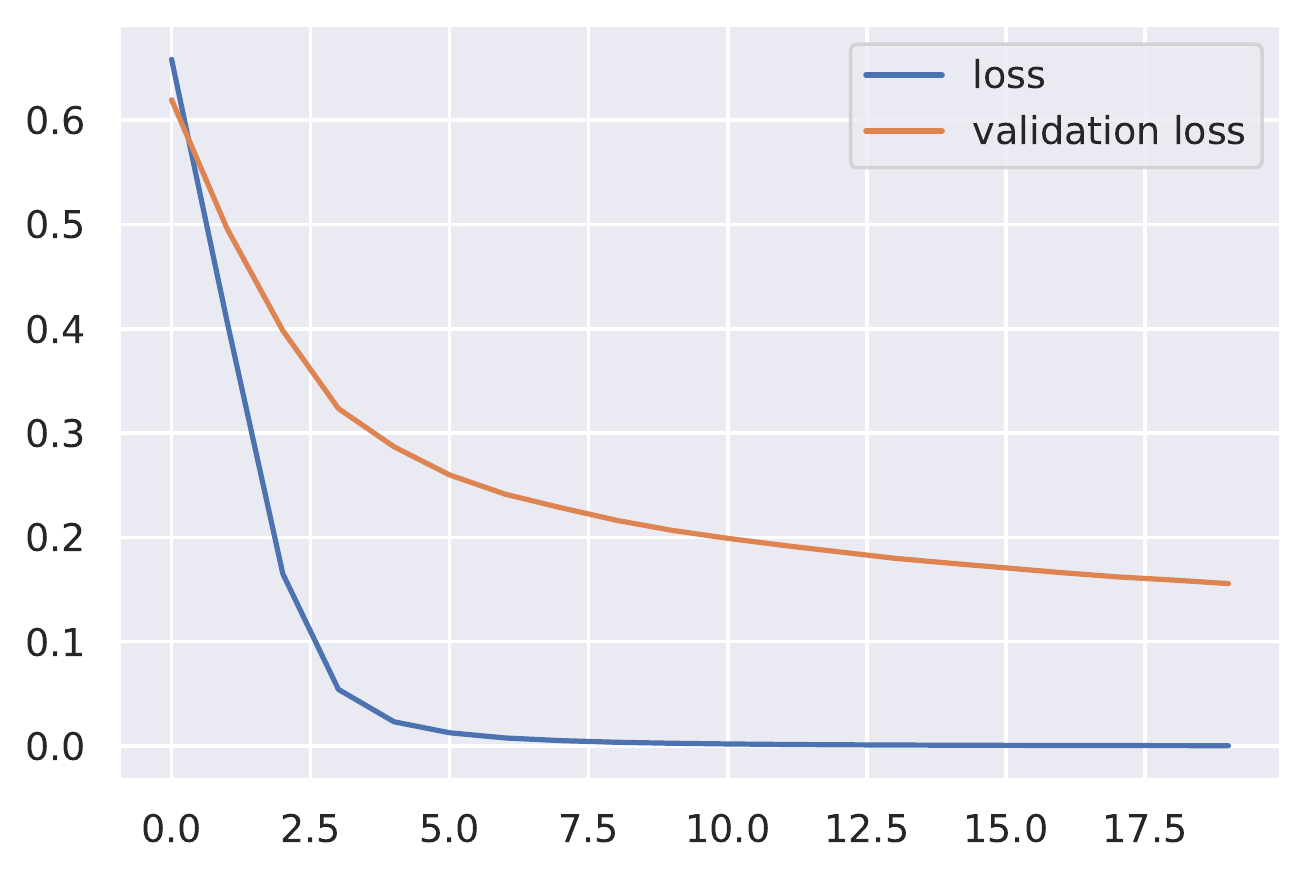}}
	\caption{Loss function vs epochs during training}
\end{figure}

The input layer of the neural network has $2^{2N+1} = 2^{13}$ (for $N=6$ qubits) nodes, which are fed the complex matrix elements of the unitary matrix flattened into an array of $2^{2N+1}$ elements. The network has two hidden layers. The first one has twice the number of nodes as the input layer ($2^{14}$) and the subsequent hidden layer has $2^9$ nodes while the output layer has a single node. All layers use ReLu \cite{relu} ($f(x) = x, \forall x>0; f(x) = 0, \forall  x<0)$ activation except for the final layer which uses a Sigmoid activation. Binary cross-entropy is a popular choice of loss function for binary classification and was the chosen loss function for the Neural Network. When trained, the output of the single output layer node will indicate if the matrix is of the type found by the algorithm or not. The actual value of the output is the probability that it is of the right type of unitary given the inputs. If the  output is greater than $0.5$ the neural net prediction is taken as that it is of the right matrix for the period finding algorithm and if lesser than $0.5$ is a prediction that it is not. Output values closer to 1  indicate a high probability  that it is of the period finding form and closer to 0 indicating it is not. The Neural network was trained and within 20 training epochs an accuracy of 100\% on the training data set was obtained with an accuracy of 95\% on the validation data set. After training the neural network, when evaluated on the test data set, showed an accuracy of  95\%. The neural network also classified  the inverse Quantum Fourier Transform matrix as of the period finding form with high degree of confidence. The output of the neural network for the inverse Quantum Fourier transform was 0.98 

\begin{figure}[ht!]
    \resizebox{8.5cm}{5cm}{\includegraphics{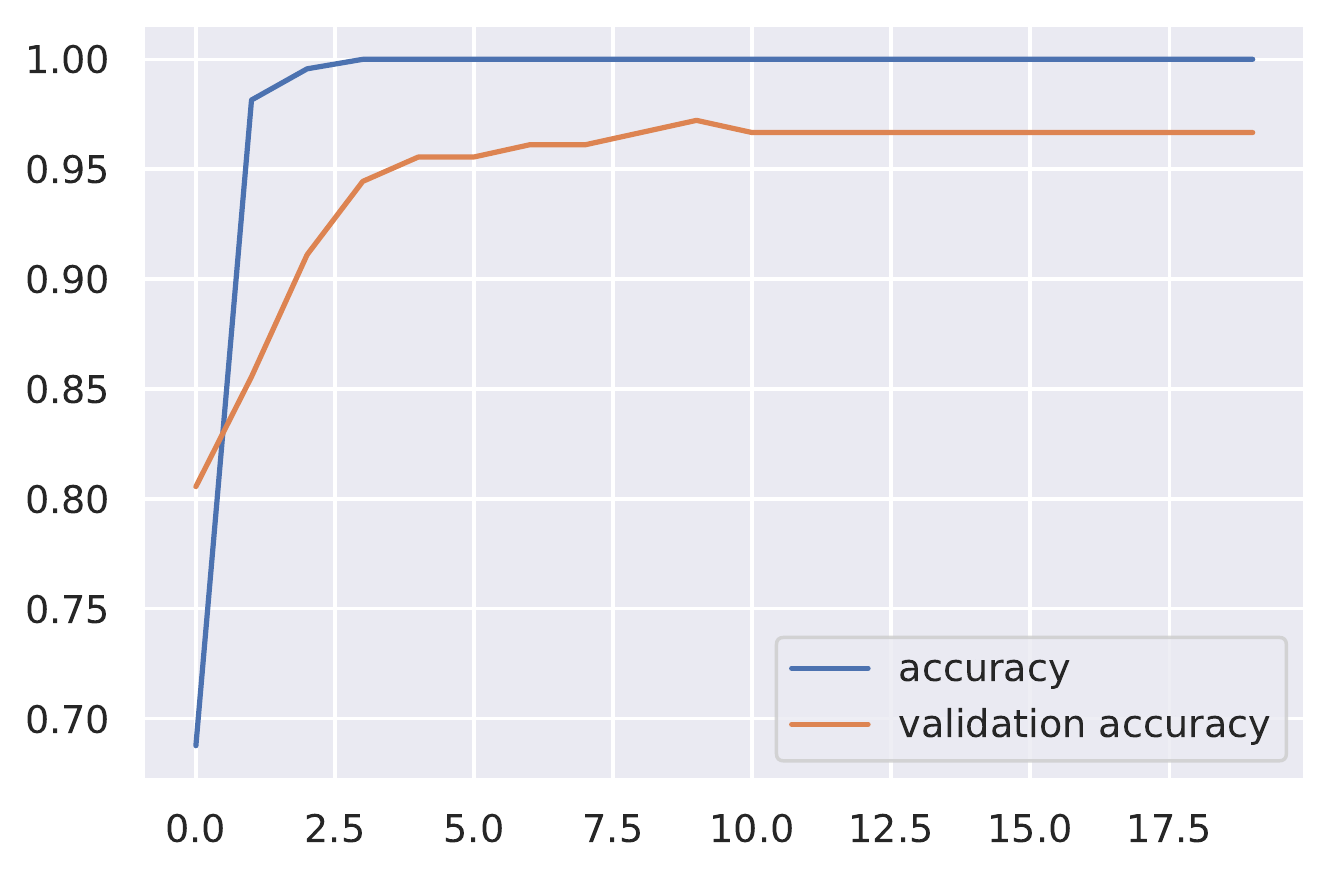}}
	\caption{Accuracy vs epochs during training}
\end{figure}

\section{Training Distribution problem \label{training}}

The desired distribution that the program trains against is an important point of consideration. It was a key piece in the design on the machine learning algorithm and in particular for formulating the cost function used in gradient descent optimization. Here the target distribution used was the one that resulted from using the inverse quantum Fourier transform as post processing unitary. This was possible in the present case because one known post-processing unitary existed. Given a problem for which one such solution is not known like the general hidden subgroup problem, the target distribution against which the machine learning algorithm can be trained is a question that has to be addressed.

In general, the question of the distribution to be trained against depends on how we can extract the required information from the distribution. A trivial probability distribution for period finding from which the period is easily apparent would be one with a single peak at $r$ for a function of period $r$. Extracting the period from this output  state would be as simple as a single measurement followed by read-out of the measured value. The machine learning algorithm, when tasked to find a unitary $M_3$ that gives this distribution, failed to reduce the cost function sufficiently close to 0 even after multiple attempts. Other choices of target distributions that were tried included a step function at $r$ and a Gaussian distribution centered around $r$ truncated at domain end points, both of which too failed to converge with cost function sufficiently close to 0. However when the output distributions of various functions corresponding to a randomly chosen post-processing unitary was fed as the target distributions to the program, it does converge to find a unitary matrix that gives the same distributions. This unitary matrix is again not the same as the original one that generated the training data. This again hints that the earlier distributions (peak at $r$, step function at $r$ and Gaussian centered around $r$) are not attainable from a unitary transformation $M_3$ 

It is possible that there is no unitary transformation that can evolve the state after oracle operation into a state with such a simple output probability distribution. The other possibility is that looking for such a unitary might by this method might be computationally infeasible. Adding ancilla qubits was another strategy that was employed to see if such a `pointer' type distribution can arise out of post-processing the oracle output. The entire system including the ancilla qubits was allowed to evolve unitarily aiming to get the desired distribution when the system qubits are subsequently measured. Adding one ancilla qubit to a 5 qubit system did not show any significant decrease in convergence of loss function to 0. Adding more ancilla qubits makes the computation intractable on available computers. The problem of identifying a target distribution that arises from a state that is unitarily connected to the oracle output state in the cases where there is no known post-processing unitary remains one that requires further investigation.  

\section{Conclusion \label{conclusion}}
Our findings suggest the unitary matrix for post processing the state after the oracle in the period finding algorithm is not unique. The machine learning program converged upon a different unitary matrix every time it was run. This is similar to the results obtained by \cite{Bang_2014} who found non uniqueness of the post oracle unitary in the Deutsch-Josza algorithm. The unitary matrices are in general significantly different. A plot of the distribution of their eigenphases however does not show any interesting features. In fact it is similar to the distribution of eigenphases of randomly generated unitary matrices. The action on $\ket{0}$ state of each of the unitary matrices generated by our program is in general different as shown by the Loschmidt echo with respect to this state. On the other hand, all these unitary matrices have a similar action on the uniform superposition state $H^{\otimes n}\ket{0}$ which is the state the oracle acts upon. 

Different unitary matrices that can implement the same quantum computation are useful with  regard to their ease of implementation on real physical quantum hardware. Certain unitary operators might be implementable with fewer gates than others in a particular hardware context. We also found that there exists a hidden pattern in the unitary matrices which was discernible to a neural network when trained on a data set of unitary matrices generated randomly and those converged on by the program. These numerical experiments show capabilities of simple feed-forward neural networks in identifying operators that can effect out the required transformation of states. The fact that the mechanism of neural-networks can be used in the domain of quantum operators is promising. For instance, it is suggestive of their applications  to a more practical use of using generative networks \cite{gan} to design operators that can effect out required transformations.  

\acknowledgements
A.S. acknowledges the support of DST-SERB through grant no. EMR/2016/007221 and the QuEST program of DST through project No. Q113 under Theme 4.

\bibliography{bibliography}
\appendix
\section{ADAM optimizer}

The function to be minimized is $f(\vec{w})$. The algorithm requires other parameters $\alpha$, $\beta_1$, $\beta_2$  and an $\epsilon$(to prevent division by zero error). At every iteration the parameters $\vec{w}$ are updated by the rules prescribed. Below is the algorithm lifted from the article\cite{adam} published by the authors of the algorithm. 

All operations on vectors are element-wise. $\vec{g_t}^2$ is the element-wise square of $\vec {g_t}$ that is  $\vec{g_t} \cdot \vec{g_t}$

\begin{algorithmic}
	\REQUIRE $\alpha$
	\REQUIRE $\beta_1, \beta_2 \in \left[ 0,1 \right)$
	\REQUIRE $\vec{w}_0 \gets $ \ random \ initial\ parameter\ vector
	\REQUIRE $f(\vec w)$, function to be minimized 
	\STATE $m_0 \gets 0$
	\STATE $v_0 \gets 0$
	\STATE $t \gets 0$

	\WHILE { $\vec{w}_t$ not converged}
	\STATE $t \gets t+1$
	\STATE $\vec g_t \gets \grad_{\vec{w}} f ( \vec w_{t-1} ) $ 
	\STATE $\vec m_t \gets \beta_1  \vec m_{t-1} + (1 - \beta_1)  \vec g_t $ 
	\STATE $\vec v_t \gets \beta_2  \vec v_{t-1} + (1-\beta_2)  \vec{g_t}^2 $
	\STATE $\hat {m_t} \gets \vec m_t / ( 1 - \beta_1^t) $ 
	\STATE $\hat {v_t} \gets \vec v_t / (1 - \beta_2^t) $ 
	\STATE $\vec{w}_t \gets \vec{w}_{t-1} - \alpha \hat{m_t} / ( \sqrt{\hat{v_t}} + \epsilon) $
	\RETURN $\vec w_t$
	\ENDWHILE
\end{algorithmic}

Four parameters $\alpha$, $\beta_1$, $\beta_2$ and $\epsilon$ are set. Vectors $\vec{m}_0$ and $\vec{v}_0$ are initialized to $\vec{0}$. At every iteration $\vec{g}_t$ (the gradient at the time step $t$) is evaluated. Using $\vec{g}_t$,  $\vec{m}_t$ and $\vec{v}_t$ the exponential moving averages of the first moment and the second moment of the gradient respectively  are calculated, the parameters $\beta_1$ and $beta_2$ control the exponential decay rates of the moving averages. These moving averages are estimates of the first moment and second moments of the gradient. Since the vectors $\vec{m}$ and $\vec{v}$ are initialized to 0 there is an initial bias in these estimates and they are corrected by dividing by $(1- \beta_1^t)$ and $(1 - \beta_2^t)$ respectively to give bias corrected moving averages $\hat{m}$ and $\hat{v}$, the article \cite{adam} describes further details regarding the bias correction. Finally the parameters of the function for the current time step $\vec{w}_t$ are updated by $\vec{w}_{t-1} + \alpha \hat{m}_t / (\sqrt(\hat{v} + \epsilon)$. The maximum step size for the update is shown to be $\alpha$, $\epsilon$ is a parameter to prevent division by zero errors

\end{document}